\documentclass[11pt,a4paper]{article}
\usepackage{amssymb}
\usepackage{graphicx}

\input{tcilatex}

\begin{document}

\title{Generalized evolutionary equations with imposed symmetries}
\author{Radu Constantinescu, Rodica Cimpoiasu \\
%EndAName
University of Craiova, 13 A.I.Cuza, 200585 Craiova, Romania}
\date{}
\maketitle

\begin{abstract}
The paper intends to propose an algorithm which could identify a general
class of pdes describing dynamical systems with similar symmetries. The way
that will be followed starts from a given group of symmetries, the
determination of the invariants and, then, of the compatible equations of
evolution. The algorithm will be exemplified by two classes of equations
which describe the Fokker-Planck model and the "backward"\ Kolgomorov one.
\end{abstract}

\section{Introduction}

Lie theory of symmetry groups for differential equations is one of the most
important methods for studying nonlinear problems which appear in physics or
in other fields of applied mathematics. The main idea of Lie's theory is the
investigation of integrability starting from the invariance of the equation
under some linear transformations of independent and dependent variables,
transformations which define the so called Lie group of symmetries. There
are different types of symmetries which can be identified for different
differential equations: point-like symmetries, contact symmetries,
classical, generalized or non-local symmetries. Many studies have been
devoted to this topic, some of them being \cite{Ibrag, Bau, Blu}. The
existence of the Lie symmetry generators for differential equations often
allows the reduction of those equations to simpler ones. The similarity
reduction method for example may be a way to follow for transforming partial
derivative equations (pdes) defined in $2D$ space in ordinary differential
equations (odes).

Usually, the direct symmetry problem of evolutionary equations is
considered. It consists in determining the symmetries of a given
evolutionary equation. The aim of this paper is to investigate the inverse
problem: what is the largest class of evolutionary equations which are
equivalent from the point of view of their symmetries. We propose an
algorithm which allows finding a general class of pdes describing dynamical
systems with similar symmetries. The problem will be effectively formulated
in the section 2 of the paper for an equation of the form $%
u_{t}(x,t)=F(t,x,u,u_{x},u_{2x})$. A general algorithm for identifying the
most general equation with a given symmetry group is presented. It is also
pointed out how this pde defined in $(1+1)$ dimensions can be reduced to an
equivalent ode by using the similarity reduction procedure. The general
results will be particularized in the section 3 of the paper for two
important models represented by the Fokker-Planck equation and "backward"\
Kolgomorov one. \ Generalized Fokker-Planck equations can be derived from
generalized linear non-equilibrium thermodynamics \cite{as2}. It was already
shown that the Fokker-Planck equation in $2D$ could be transformed, by
appropiate change of coordinates, in heat equation or Schrodinger equation 
\cite{as1}. As far as Kolmogorov equation is concerned, it has been
considered even in infinite dimensions and interesting applications to
stochastic generalized Burgers equations have been noted \cite{as5}. Some
concluding remarks will end the paper.

\section{The general setting of the problem}

Let us consider a general dynamical system described in a $(1+1)$
dimensional space $(x,t)$ by a second order partial derivative equation of
the form:%
\begin{equation}
u_{t}=A(x,t,u)u_{2x}+B(x,t,u)u_{x}+C(x,t,u)u,\text{ }A\neq 0  \label{1}
\end{equation}%
The coefficient functions $A(x,t,u),~B(x,t,u),~C(x,t,u)$ are arbitrary $C^{1}
$ functions. Any equation $\Delta _{1}$ of the form (\ref{1}) is invariant
under the infinitesimal transformations:

\begin{equation}
\bar{x}=x+\xi (t,x,u)\varepsilon +...;~\bar{t}=t+\varphi (t,x,u)\varepsilon
+...;\ \bar{u}=t+\eta (t,x,u)\varepsilon +...  \nonumber
\end{equation}%
if and only if it satisfies \cite{Olver}:

\begin{equation}
U^{(n)}(\Delta _{1})\mid _{\Delta _{1}=0}=0  \label{1b}
\end{equation}%
where $U^{(n)}$ is the $n-$th extension of the symmetry operator $U$. In the
case of the equation (\ref{1}) defined in $(1+1)$ dimensions, the operator $%
U $, also known as Lie operator, has the general form:

\begin{equation}
U(x,t,u)=\varphi (x,t,u)\frac{\partial }{\partial t}+\xi (x,t,u)\frac{%
\partial }{\partial x}+\eta (x,t,u)\frac{\partial }{\partial u}  \label{2}
\end{equation}%
As in our case the equation (\ref{1}) is a second order differential
equation, the invariance condition (\ref{1b}) will be in fact:%
\begin{equation}
U^{(2)}[u_{t}-A(x,t,u)u_{2x}-B(x,t,u)u_{x}-C(x,t,u)u]=0  \label{3}
\end{equation}%
where $U^{(2)}$ is the second order extension of the Lie symmetry operator $%
U $. A concrete computation leads to the equation:

\begin{eqnarray}
0 &=&(\varphi A_{t}+\xi A_{x}+\eta A_{u})u_{2x}+(\varphi B_{t}+\xi
B_{x}+\eta B_{u})u_{x}+\varphi uC_{t}+\xi C_{x}u+  \label{4} \\
&&+C\eta +\eta C_{u}u+B\eta ^{x}-\eta ^{t}+A\eta ^{2x}  \nonumber
\end{eqnarray}%
By substituting in (\ref{4}) the general expressions given in \cite{Olver}
for $\eta ^{x},$ $\eta ^{t},$ $\eta ^{2x}$ and asking for the vanishing of
the coefficients of each monomial in the derivatives of $u(t,x)$, one
obtains the following differential system:%
\[
\varphi _{x}=0;\ \varphi _{u}=0;\ \xi _{u}=0;\ \eta _{2u}=0;-\varphi
A_{t}-\xi A_{x}-\eta A_{u}-A\varphi _{t}+2A\xi _{x}=0; 
\]%
\begin{equation}
-\varphi B_{t}-\xi B_{x}-\eta B_{u}+B\xi _{x}-\xi _{t}-B\varphi _{t}-2A\eta
_{xu}+A\xi _{2x}=0  \label{5a}
\end{equation}%
\[
-\varphi C_{t}u-\xi C_{x}u-C\eta -\eta C_{u}u-B\eta _{x}+\eta _{t}+C\eta
_{u}u-\varphi _{t}Cu-A\eta _{2x}=0 
\]%
\bigskip

It is important to note that this system can be seen as a general system
with the unknown functions $A(x,t,u),$ $B(x,t,u),$ $C(x,t,u),$ $\varphi
(x,t,u),$ $\xi (x,t,u),$ $\eta (x,t,u).$Two completely different situations
can be considered:

- one fix the equation (\ref{1}) by choosing concrete expressions for $%
A(x,t,u),$ $B(x,t,u),$ $C(x,t,u)$ and one determines the form of the
symmetries, that is $\varphi (x,t,u),$ $\xi (x,t,u),$ $\eta (x,t,u);$

- one looks for the class of equations which have a given form of the
symmetry, which means to solve (\ref{5a}) with given $\varphi (x,t,u),$ $\xi
(x,t,u),$ $\eta (x,t,u)$ for the unknown functions $A(x,t,u),$ $B(x,t,u),$ $%
C(x,t,u)$.

Usually the first approach is considered. What we are doing here is to
perform the second approach. We will start from the already known symmetries
of the Fokker-Planck equation \cite{as2}, an equation which belongs to the
general class of equations (\ref{1}) and describes various evolutionary
processes in quantum optics, solid state physics or statistical physics. Let
us therefore consider for the Lie symmetry operator (\ref{2}) the following
form:%
\begin{equation}
U(x,t,u)=e^{2t}\frac{\partial }{\partial t}+xe^{2t}\frac{\partial }{\partial
x}-x^{2}e^{2t}u\frac{\partial }{\partial u}  \label{6}
\end{equation}%
With these choices for $\varphi (x,t,u),$ $\xi (x,t,u),$ $\eta (x,t,u)$, the
system (\ref{5a}) reduces to the equations:%
\begin{eqnarray}
A_{t}+xA_{x}-x^{2}A_{u}u &=&0  \nonumber \\
-B_{t}-xB_{x}+x^{2}uB_{u}-B-2x+4Ax &=&0  \label{7} \\
C_{t}u+C_{x}ux-C_{u}u^{2}x^{2}-2Bxu+2x^{2}u+2Cu-2Au &=&0  \nonumber
\end{eqnarray}%
It is a system with the unknown functions $A(x,t,u),$ $B(x,t,u),$ $C(x,t,u)$
which leads to the general solution of the form:%
\begin{eqnarray}
A &=&f\left( t-\ln x,ue^{x^{2}/2}\right) ,  \nonumber \\
B &=&\left[ -1+2f\left( t-\ln x,ue^{x^{2}/2}\right) \right] x+\frac{g\left(
t-\ln x,ue^{x^{2}/2}\right) }{x},\text{ }  \label{8} \\
C &=&\left[ -1+f\left( t-\ln x,ue^{x^{2}/2}\right) \right] x^{2}+f\left(
t-\ln x,ue^{x^{2}/2}\right) +g\left( t-\ln x,ue^{x^{2}/2}\right) +  \nonumber
\\
&&+\frac{h\left( t-\ln x,ue^{x^{2}/2}\right) }{x^{2}}.  \nonumber
\end{eqnarray}%
The solution (\ref{8}) is given in terms of 3 arbitrary functions $f(u,x,t),$
$g(u,x,t),$ $h(u,x,t)$. In fact, two types of combinations of the variables $%
(u,x,t)$ have to be considered: 
\begin{equation}
\widetilde{z}\equiv (t-\ln x),\phi (z)\equiv ue^{x^{2}/2}  \label{8b}
\end{equation}%
The significance of these variables becomes clear if one considers the
characteristic equations associated with the symmetry operator (\ref{6}):%
\begin{equation}
\frac{dt}{e^{2t}}=\frac{dx}{xe^{2t}}=\frac{du}{-x^{2}e^{2t}u}  \label{9}
\end{equation}%
By integrating the equations (\ref{9}) the following similarity variables
are obtained:%
\begin{equation}
z=xe^{-t},\phi (z)=ue^{x^{2}/2}  \label{10}
\end{equation}%
It is simple now to notice the direct connection between (\ref{8b}) and (\ref%
{10}), that is the general solution (\ref{8}) is given in terms of
similarity variables. Moreover, it is also important to note that, by using
the similarity transformations (\ref{10}), the class of the evolutionary
systems which are described by the $(1+1)$ equation (\ref{1}) can be reduced
to the $1D$ systems with the evolution given by the following differential
equation:

\begin{equation}
z^{2}\frac{d^{2}\phi (z)}{dz^{2}}f\text{ }\left[ \phi (z),-\ln z\right] +z%
\frac{d\phi (z)}{dz}g\text{ }\left[ \phi (z),-\ln z\right] +\phi (z)h\left[
\phi (z),-\ln z\right] =0,\text{ }\forall \text{ }f,g,h  \label{11}
\end{equation}%
As a matter of fact, let us note that for $f=1,$ $g=h=0,$ the equation (\ref%
{11}) becomes:%
\begin{equation}
\frac{d^{2}\phi (z)}{dz^{2}}=0  \label{15}
\end{equation}%
One obtains a wave type equation in one dimension with the solution of the
form:%
\begin{equation}
\phi (z)=c_{1}z+c_{2}  \label{15'}
\end{equation}%
where $c_{1},c_{2}$ are arbitrary constants.

As a conclusion of this section, let us mention that the solution (\ref{8})
gives the most general class of $2D$ equations of the form (\ref{1}) which
admits a symmetry operator of the form (\ref{6}) and, by that, accepts a
similarity reduction to the $1D$ equation (\ref{11}). In the next section we
will identify two concrete examples of systems which belong to this class of
evolutionary equations.

\section{Applications}

\subsection{The Fokker-Planck model}

Let us come back to the choice $f=1,$ $g=h=0$ in the equation (\ref{11}),
choice already considered before. We have to note that for this case the
solution (\ref{8}) of the system (\ref{7}) transforms into very simple
expressions:%
\begin{equation}
A=1,\text{ }B=x,\text{ }C=1  \label{12}
\end{equation}%
The equation (\ref{1}) takes, it too, a very simple form known as the $(1+1)$
Fokker-Planck equation \cite{Planck}:%
\begin{equation}
u_{t}=u_{2x}+xu_{x}+u  \label{14}
\end{equation}%
One recovers so that the equation (\ref{14}) is one of the particular form
of the equation (\ref{1}) which accepts the symmetry (\ref{6}). As we
mentioned, this is an already known result and, in fact, it justifies the
choice of the symmetry operator in the form (\ref{6}). The equation (\ref{14}%
) is used in various fields of natural sciences, as physical chemistry or
theoretical biology \cite{FK}. In the similarity variables (\ref{10}) it can
be reduced to form (\ref{15}) and can be solved. Coming back to the original
variables $x,t$ and $u=u(x,t)$ by inverting (\ref{10}), the solution (\ref%
{15'}) of the equation becomes:

\begin{equation}
u(x,t)=c_{1}xe^{-t}e^{-x^{2}/2}+c_{2}e^{-x^{2}/2}  \label{16}
\end{equation}%
So, it was possible to obtain a solution of the equation (\ref{16}) as a
simple case of the general results obtained for the equation (\ref{1}). We
have to mention that, by computational methods, one can obtain a general
solution of the equation (\ref{16}) in a very complicated form. Such results
are mentioned in \cite{as3}, \cite{as4}.

\subsection{The "backward" Kolmogorov model}

Let us consider now another particular case of solution in (\ref{8}): $f=1,$ 
$g=-1,h=0$. In this case the equation (\ref{1}) takes the form of an
evolutionary equation of the type "backward" Kolmogorov \cite{Kol}:%
\begin{equation}
u_{t}=u_{2x}+(x-\frac{1}{x})u_{x}  \label{18}
\end{equation}%
As the whole class of equations belonging to (\ref{1}) and as the
Fokker-Planck equation, the equation (\ref{18}) accepts the Lie symmetry (%
\ref{6}) and can be reduced to a one-dimensional ordinary differential
equation of the form:%
\begin{equation}
z\frac{d^{2}\phi (z)}{dz^{2}}-\frac{d\phi (z)}{dz}=0  \label{19}
\end{equation}%
The equation (\ref{19}) has a solution of the form:%
\begin{equation}
\phi (z)=c_{3}+c_{4}z^{2}  \label{20}
\end{equation}%
Using again (\ref{10}) one transforms (\ref{20}) into a solution of (\ref{18}%
) of the form: 
\begin{equation}
u(x,t)=e^{-x^{2}/2}(c_{3}+c_{4}x^{2}e^{-2t})  \label{22}
\end{equation}

\section{Concluding remarks}

The results of this paper can be synthesized as follow: $(i)$ the class of $%
(1+1)$ generalized evolutionary equations of type (\ref{1}) can allow the
same concrete Lie symmetry operator (\ref{6}) if and only if the coefficient
functions $A(x,t,u),B(x,t,u),C(x,t,u)$ have the expressions (\ref{8}); $(ii)$
the knowledge of symmetry operators enabled to introduce, by integrating the
characteristic equations (\ref{9}), the so called similarity variables.
Through this similarity approach, the general form for reduced odes (\ref{11}%
) associated to the analyzed equations (\ref{1}) has been obtained; $(iii)$
the results are particularized for two dynamical equations, the
Fokker-Planck model and the "backward"\ Kolgomorov one. The general
algorithm allows deriving simple solutions of the evolutionary equations
which describe the two models.

\textbf{Acknowledgements}: The paper was supported by the Romanian Ministry
of Education, Research and Innovation, by the grant "Ideas" 2008, code ID
418.


\bigskip

\begin{thebibliography}{99}
\bibitem{Ibrag} N.H. Ibragimov, ,,Handbook of Lie Group Analysis of
Differential Equations\textquotedblright , Volume1,2,3 CRC Press, Boca
Raton, Ann Arbor, London, Tokyo, (1994,1995,1996).

\bibitem{Bau} G.Baumann, ,,Symmetry Analysis of Differential Equations with
Mathematica\textquotedblright , Telos, Springer Verlag, New York (2000).

\bibitem{Blu} G.W. Bluman, S.C.Anco, ,,Symmetry and Integration Methods for
Differential Equations\textquotedblright , Springer-Verlag, New York (2002).

\bibitem{as2} T.D. Frank, Physica A 310 (2002), 397-412.

\bibitem{as1} S. Spichak and V. Stognii, Proceedings of Institute of
Mathematics of NAS of Ukraine, Vol. 30, Part 1, (2000), 204--209.

\bibitem{as5} M. Rockner, Z. Sobol, C. R. Math. Acad. Sci. Paris, Ser. I 338
no. 12, (2004), 945-949.

\bibitem{Olver} P. J.Olver, ,,Applications of Lie Groups to Differential
Equations\textquotedblright , GTM 107, Second edn., Springer-Verlag (1993).

\bibitem{Planck} G. I. Burde, Proceedings of Institute of Mathematics of NAS
of Ukraine,Vol. 43, Part 1, (2002), 93--101

\bibitem{FK} H. Risken, ,,The Fokker--Planck Equation: Methods of Solutions
and Applications\textquotedblright , 2nd edition, Springer Series in
Synergetics, Springer, ISBN 3-540-61530-X, (1989).

\bibitem{as3} S. F. Wojtkiewicz , L.A. Bergman, Proceedings of the 8th
International Conference on Structural Safety and Reliability, Newport
Beach, (2001).

\bibitem{as4} S. F. Wojtkiewicz, L.A. Bergman , B.F. Spencer Jr., E. A.
Johnson, Nonlinear and Stochastic Structural Dynamics, (2001), 277-288.

\bibitem{Kol} J.Biazar, Int. Math. Forum, 3, no 19, (2008), 945-954.
\end{thebibliography}
\end{document}